\documentclass[twocolumn]{jpsj2} 
\newcommand{\G}{\Gamma}
\newcommand{\dg}{\dagger}

\title{Local Heavy Quasiparticle in Four-Level Kondo Model}

\author{Kazumasa \textsc{HATTORI}, Yosuke \textsc{HIRAYAMA} and Kazumasa \textsc{MIYAKE}}

\inst{Division of Materials Physics, Department of Materials Engineering Science, Graduate School of Engineering Science, Osaka University, Toyonaka, Osaka 560-8531, Japan}

\abst{An impurity four-level Kondo model, in which an ion is tunneling among 4-stable points and interacting with surrounding conduction electrons, is investigated using both perturbative and numerical renormalization group methods. The results of numerical renormalization group studies show that it is possible to construct the ground state wavefunction including the excited ion states if we take into account the interaction between the conduction electrons and the ion. The resultant effective mass of quasiparticles is moderately enhanced. This result offers a good explanation for the enhanced and magnetically robust Sommerfeld coefficient observed in SmOs$_4$Sb$_{12}$, some other filled-skutterudites,  and clathrate compounds.}

\kword{renormalization group, numerical renormalization group, heavy fermion, Kondo effect, rattling}

\begin{document}
\maketitle
\section{Introduction}
 Since the two-level Kondo effect was studied about a three decades ago,\cite{Kondo1,Kondo2} the physics arising from the off-center degrees of freedom of ions have attracted much attention.
 Recently, off-center motions of ions in clathrate compounds\cite{NemotoCePd} and  filled-skutterudites\cite{GotoPrOs} are identified by ultrasonics. A dispersion of the sound velocity and a softening in the elastic constants are observed in various materials such as PrOs$_4$Sb$_{12}$, LaOs$_4$Sb$_{12}$, Ce$_3$Pd$_{20}$Ge$_6$, and so on. Characteristic properties of these materials are the existence of a thermally excited rattling motion in the ``cage'', and a possible {\it realization of an anisotropic ground state with respect to the ion configurations}. For example, the elastic constant $C_{44}$ of La$_3$Pd$_{20}$Ge$_6$\cite{RevClath} displays a softening at least down to 20mK. This implies the existence of a triply degenerate mode in the ground state not in the singlet state which is expected in a simple quantum mechanical system. 

About two decades ago, the energy level scheme for a simple and canonical off-center mode was obtained in NaCl including OH$^-$ in its interstitials\cite{NaCl,NaCl2}. The OH$^-$ ion is located at the [100] off-center positions in NaCl. The result of the absorption spectrum of a millimeter wave supports a singlet ground state ($\G_1^+$) with an excited state triplet ($\G_4^{+}$). In that case, no softenings in the elastic constants is observed at low temperatures, reflecting the singlet ground state. 

The off-center positions in PrOs$_4$Sb$_{12}$ and LaOs$_4$Sb$_{12}$ are expected to be the same. In the case of PrOs$_4$Sb$_{12}$, in addition to the f-electron contributions, the elastic constant $C_{11}-C_{12}$ continues to display an anomalous softening to the temperature just above the superconducting transition temperature\cite{GotoPrOs}. This implies that, in a low temperature region, there exists a $\Gamma_3 (\Gamma_{23})$ mode of the ion configurations, while an elementary quantum mechanics tells us that the lowest energy state is the singlet with a symmetric ion configuration in such tunneling systems. In this point, the observed results are mysterious. It is natural to expect that a magnetically robust heavy fermion compound SmOs$_4$Sb$_{12}$\cite{Sm} is added into such a group of materials.

As theoretical developments for the two-level systems (TLS), it was pointed out that a strong interaction between local phonons and conduction electrons generates such off-center situations as a model of A15 compounds\cite{YuAnderson,MatsuMk,KusuMyk}. Recently these problems are also discussed in the lattice model in relation to the rattling degrees of freedom in filled-skutterudite compounds\cite{MikimotoOno}. The problems of the off-center motions of ions in the two-level system coupled with conduction electrons were studied extensively\cite{VladZw1,VladZw2,VladZw3,ZarandZwadow,Moustakas1,Moustakas2,Ye} in the context of the two-channel Kondo model.

The purpose of this paper is to study a property of a four-level Kondo model which is a simplified model for the filled-skutterudite compounds with apparent off-center degrees of freedom of ions in the cage. In particular, a condition for an appearance of heavy quasiparticles is investigated on the basis of not only the perturbative renormalization group (PRG) method but also the numerical renormalization group (NRG) one.

The systems with off-center positions more than two were also discussed so far. $M$-level system was discussed in ref. 20 and a symmetric SU($M$)$\otimes$SU($N_f$) Coqblin-Schrieffer type effective Hamiltonian was derived using $1/N_f$ expansion, where $N_f$ is a number of the conduction electron channels (spin). In Ref. 20, the only conduction electron orbital which couples most strongly to the ion is retained. In this paper we investigate the properties, which arise from the purely orbital degrees of freedom of the conduction electrons.

This paper is organized as follows: In \S\ 2, we derive the effective model Hamiltonian for a four-level Kondo model. The details of explicit derivation of four-level Kondo model are given in the Appendix A.  In \S\ 3, we discuss the property of the four-level Kondo model using scaling approach on the basis of PRG. In \S\ 4, we discuss the four-level Kondo model using NRG method\cite{Wilson}. In \S\ 5, we give a possible explanation for the unusual properties observed in the above-mentioned materials. Finally, \S\ 6 summarizes the result of the paper.

\section{Models}
In this section, we derive an effective model Hamiltonian describing the four-level off-center mode coupled with conduction electrons. 
\subsection{Effective model Hamiltonian}
\begin{figure}[t]
	\begin{center}
    \includegraphics[width=.4\textwidth]{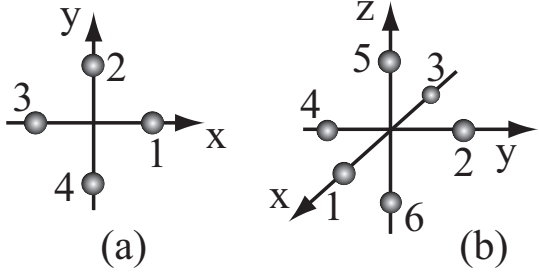}
  \end{center}
\caption{$(a,0,0)$ type off-center positions of an ion. Each circle represents off-center positions located at $(\pm a,0,0)$, $(0,\pm a,0)$ and $(0,0,\pm a)$. The numbers correspond to indices of $\phi_i({\bf x})$. (a)two-dimensional square case. (b) three-dimensional cubic case.}
\label{fig-1}
\end{figure}

First, we assume that the ground-state wavefunction of an ion localized at each off-center position ${\bf x}_i \ \ (i=1,2,3,\cdots)$ is approximated by the Gaussian form as,
\begin{eqnarray}
 \phi_i^0({\bf x})= \Big(\frac{1}{\pi\sigma}\Big)^{\frac{3}{4}}\exp\Big[-\frac{({\bf x}-{\bf x}_i)^2}{2\sigma^2}\Big],
\end{eqnarray}
where $\bf x$ is the coordinate of the ion, $\sigma$ is a broadening length of ion wavefunction. In this paper, we investigate the case  where the off-center positions are $(\pm a,0)$, $(0,\pm a)$, see Fig.\ref{fig-1}(a). The state $|i\rangle$ corresponding to $\phi_i^0({\bf x})$ is mixed with $|j\rangle\ (i\not=j)$ by the quantum mechanical tunneling. Note that we consider the low temperature physics so that thermally excited classical tunneling process are discarded, or such processes are renormalized in the parameters of the low temperature effective Hamiltonian.
The overlap $S_{ij}$ between states $|i\rangle$ and $|j\rangle$ are estimated by an usual integral as follows:
\begin{eqnarray}
 S_{ij}\!\!\!\!&=&\!\!\!\! \int d^3x \ {\phi_i^0}^{\ast}({\bf x}){\phi_j^0}({\bf x})\\
\!\!\!\!&=&\!\!\!\!{\Bigg\{ }
\begin{array}{@{\,}c@{\,}}
    e^{-\frac{a^2}{2\sigma^2}} \equiv S\ll1\ \ \rm for\ the\ nearest\ neighbors,\\
    S^2 \ \ \ \ {\rm for\ the\ {\rm next-}nearest\ neighbors.}
\end{array}
\label{S}
\end{eqnarray}
Because of property (\ref{S}) we retain only the nearest-neighbor hopping for simplicity. These tunneling terms cause the energy splitting with respect to the irreducible representations of ion wavefunctions, as discussed in the next section.

Next, we consider conduction-electron assisted tunneling. These processes are occurred by the following matrix elements,

\begin{eqnarray}
\frac{1}{\Omega}\int\!\!\!\!\!\int \!\!d^3x d^3r {\phi_i^0}^{\ast}({\bf x})e^{-{\rm i}\bf k\cdot r}U({\bf x-r})e^{{\rm i}{\bf k'\cdot r}}{\phi_j^0}({\bf x})\\
= F^0_{ij}({\bf k}-{\bf k}')U_{{\bf k}-{\bf k}'}.
\end{eqnarray}
Here, $U({\bf x-r})$, $U_{{\bf k}-{\bf k}'}$ and $\Omega$ are the screened Coulomb interaction between the ion and the conduction electrons, its Fourier transform and the system volume, respectively. $F^0_{ij}({\bf q})$ is given by
\begin{eqnarray}
F^0_{ij}({\bf q})&=&\int d^3xe^{-{\rm i}{\bf q \cdot x}}{\phi_i^0}^{\ast}({\bf x}){\phi_j^0}({\bf x}) \label{eqF0}\\
&=&e^{-\frac{\sigma^2|{\bf q}|^2}{4}}\exp[-\frac{({\bf x}_i-{\bf x}_j)^2}{4\sigma^2}-{\rm i}\frac{({\bf x}_i+{\bf x}_j)}{2} \cdot {\bf q} ].\nonumber\\
\label{Fij}
\end{eqnarray}

Before writing a Hamiltonian in this system, we should orthogonalize $\phi_i^0({\bf x})$.\cite{CoxZW} To leading order in $S$, we obtain the orthogonalized wavefunctions:
\begin{eqnarray}
\phi_i({\bf x})\sim \phi_i^0({\bf x})-\frac{1}{2} S \sum_j \phi_j^0({\bf x}).
\end{eqnarray}
Here, the summation of $j$ runs over the nearest-neighbor off-center positions of $i$.

From eqs. (\ref{S}) and (\ref{Fij}) the Hamiltonian describing the off-center motion of the ion and the interaction between the ion and the conduction electrons are given as
\begin{eqnarray}
H&=&\sum_{ij}\Delta_{ij}\phi_i^{\dg}\phi_{j}+\sum_{\bf k\mu}\epsilon_{{\bf k}}c_{{\bf k}\mu}^{\dg}c_{{\bf k}\mu}+H_{{\rm int}},\label{Heff}\\
H_{{\rm int}}&=&\sum_{{\bf kk'}\mu}U_{\bf k-k'}c_{{\bf k}\mu}^{\dg}c_{{\bf k'}\mu}\sum_{ij}F_{ij}({\bf k-k'})\phi_i^{\dg}\phi_{j},\label{Hint}
\end{eqnarray}
where $c_{{\bf k}\mu}$ and $\phi_i$ is the annihilation operator of the conduction electron with the momentum $\bf k$, spin $\mu$ and that of the ion (orthogonalized) with the off-center position ${\bf x}_i$, respectively. The tunneling amplitude $\Delta_{ij}$ is estimated as $\Delta_{ij}\sim -\hbar\omega_0 S_{ij}$ where $\omega_0$ is the frequency of the ion around each off-center positions. $F_{ij}({\bf k-k'})$ is estimated by using $\phi_i({\bf x})$ instead of $\phi_i^0({\bf x})$ in eq. (\ref{eqF0}). Note that an ion operator $\phi_i$ acts on a restricted Fock space with $\sum_i\phi_i^{\dag}\phi_i=1$. Non-commutative properties of this model come from the off-diagonal terms of $H_{\rm int}$, which give rise to, in simple cases, various Kondo effects.

In the following, we ignore the spin of the conduction electrons and expand $c_{{\bf k}}$ by the spherical harmonics at the origin as in the usual way of an impurity model. Physical properties generated from the spin dependence will not be important because the resulting effective Hamiltonian is highly anisotropic in the orbital space of the conduction electrons (s-, p-, d-wave), {\it e.g.}, giving little chance for the two-channel Kondo effect to be realized. Then we expect that criticalities such as the two-channel Kondo model cannot arise in the present model. Moreover, The existence of the splittings $\Delta_{ij}$ also helps this expectation.

\subsection{Order estimation of parameters}
Before proceeding to detailed analyses, we need to estimate the order of magnitude of parameters in our effective Hamiltonian eq. (\ref{Heff}). Using a one-dimensional double well potential,
\begin{eqnarray}
V(x)=-\alpha x^2+\beta x^4,
\end{eqnarray}
the stable points $\pm|x^{\ast}|$ are estimated as $|x^{\ast}|=\sqrt{\frac{\alpha}{2\beta}}\equiv \sigma w^{-1}$, where $w$ is a parameter of $O(1)$. 
We can estimate the frequency $\omega_0$ of the harmonic oscillator centered at $\pm |x^*|$ as $\hbar \omega_0\sim 8wV_B$, where $V_B$ is the height of the barrier at $x\sim 0$. For the nearest-neighbor pair $(i,j)$, $\Delta_{ij}\sim \hbar \omega_0 S=8w^2V_B S$ and $U({\bf q})F_{ij}({\bf q})\sim U({\bf q})S\sim\Delta_{ij}U({\bf q})/(\hbar\omega_0)=\Delta_{ij}U({\bf q})/(8V_Bw^2)$. These expressions coincide with the result of the detailed analysis in an order of magnitude given in ref. 21 
. In the case of PrOs$_4$Sb$_{12}$, La$_3$Pd$_{20}$Ge$_6$ and Ce$_3$Pd$_{20}$Ge$_6$, $V_B$ is estimated as $70-200$K\cite{RevClath}. Taking into account the fact, $a\sim|x^{\ast}|$, $S\sim {\exp}({-1/(2w^2)})=0.0183$ for $w=1/(2\sqrt{2})$. Then $\Delta_{ij} \simeq$1K in the case that $(i,j)$ is in the nearest neighbor with each other. An ambiguity about the interaction term $U({\bf q})$ remains. There are two limiting cases: $|U({\bf q})|\gg V_B\ (|U({\bf q})|\simeq 1000{\rm K})$  and $|U({\bf q})|\ll V_B\ (|U({\bf q})|\simeq 10{\rm K})$. Detailed investigations will be given in later sections.

\section{Perturbative Renormalization Group Approach}\label{3}
 In this section, we calculate the effective interactions using PRG method, for the purpose of seeing the overall features of the scaling in our model. 
Although there exist some studies by using renormalization group in such models\cite{MLS}, our approach is a natural extension of refs. 13 and 14. A change from refs. 13 and 14 is a treatment for bandwidth $2D_0$ of the bare conduction electrons. It was pointed out in ref. 23 that $D_0$ should be replaced by an energy level of the first excited state ($\epsilon_3$ in their notation) which is smaller than $D_0\ \ (\epsilon_3 \sim (10^{-3}\sim10^{-2})\times D_0)$. Because of this, it is concluded that the Kondo temperature is always smaller than the spontaneous splitting of the ion configuration, in {\it weak coupling}. Recently a mechanism of a resonant scattering of the tunneling impurity was proposed in order to obtain rather high Kondo temperature\cite{resonant}. In this paper, we consider the case $D_0\sim 100$K$\sim V_B$. Note that $V_B$ is comparable to $\hbar\omega_0$ which is an energy spacing of the local phonon, i.e. Einstein frequency.

\setcounter{equation}{0}
\subsection{Effective exchange Hamiltonian}
 We expand the annihilation operator $c_{\bf k}$ of the conduction electron in terms of the spherical harmonic function $Y_{lm}$ as,
\begin{eqnarray}
  c_{\bf k}= \sum_l \sum_{m=-l}^{l} Y_{lm}(\hat{k}){\rm i }^lc_{klm},\label{expLM}
\end{eqnarray}
where we define a new annihilation operator $c_{klm}$ with $k=|\bf k|$, the angular momentum $l$ and its z-component $m$. In eq. (\ref{expLM})
we have omitted a coefficient depending on k\cite{Yotsu}, which causes no significant problem by absorbing it in a definition of the bandwidth $2D_0$ (centered at the Fermi energy). The interaction $U_{\bf k-k'}$ is also expressed by the Legendre expansion as,

\begin{eqnarray}
  U_{\bf k-k'} = \sum_l \sum_{m=-l}^{l}\frac{4\pi}{2l+1} u_{l}Y_{lm}(\hat{k})Y_{lm}^{\ast}(\hat{k'}), \label{U}
\end{eqnarray}
where $u_l$ is the $l$-th coefficient of the Legendre expansion for $U_{\bf k-k'}$.
 Then Hamiltonian (\ref{Heff}) can be written as
\begin{eqnarray}
H&=&\sum_{\mu}\Delta_{\mu}a^{\dag}_{\mu}a_{\mu}+\sum_{k\hat{l}}\epsilon_{k}c^{\dag}_{k\hat{l}}c_{k\hat{l}}+H_{\rm int},\\
H_{\rm int} &=& \sum_{kk'}\sum_{\hat{l}\hat{l'}}\sum_{\mu\nu}\sum_{\gamma} J_{\hat{l}\hat{l'}}^{\gamma}c_{k\hat{l}}^{\dag}c_{k'\hat{l'}}a_{\mu}^{\dag}{\bf \Psi}^{\gamma}_{\mu\nu}a_{\nu}.\label{HintKondo}
\end{eqnarray}
 Here, we have introduced new ion annihilation operators $a_{\mu}$'s which are the irreducible representation of the point group with the energy level $\Delta_{\mu}$ of the ion configuration (\S\ 3.2). The indices $\hat{l}$ and $\hat{l'}$ represent the angular momentum $(l,m)$. The coupling constant $J_{\hat{l}\hat{l'}}^{\gamma}$ is calculated in Appendix A for the four-level system. $\Psi^{\gamma}_{\mu\nu}$ is matrices of the $\gamma$-th irreducible representation of the direct product $a^{\dag}_{\mu}a_{\nu}$ and normalized as $\sum_{\mu\nu}|\Psi^{\gamma}_{\mu\nu}|^2=1$ (\S\ 3.2).

 For this kind of rather general exchange Hamiltonian (\ref{HintKondo}), a formula of 2-loop renormalization group equations has been derived in ref. 25
. The renormalization group equation for $\Delta_{\mu}$ is obtained in a way similar to the case of TLS. Then a set of renormalization group equations is given as follows:
\begin{eqnarray}
\!\!\!\!\!\!\!\!\!\!\!\frac{\partial J_{\hat{l}\hat{l'}}^{\gamma}}{\partial x}\!\!\!\!\!&=& \!\!\!\!\!\rho\sum_{\hat{l''}}\sum_{\alpha\beta}J_{\hat{l}\hat{l''}}^{\alpha}J_{\hat{l''}\hat{l'}}^{\beta}{\rm Tr}\Big([{\bf \Psi}^{\alpha}, {\bf \Psi}^{\beta}] {{\bf \Psi}^{\gamma}}^{\dag}\Big)\nonumber\\
\!\!\!\!\!&+&\!\!\!\!\!\sum_{\alpha\beta\lambda}\frac{\rho^2}{2}{\rm Tr}\Big(J^{\alpha}J^{\beta}\Big)J_{\hat{l}\hat{l'}}^{\lambda}{\rm Tr}\Big([{\bf \Psi}^{\alpha},[{\bf \Psi}^{\beta}, {\bf \Psi}^{\lambda}]]{{\bf \Psi}^{\gamma}}^{\dag}\Big),\label{RG1}\\
\!\!\!\!\!\!\!\!\!\!\!\frac{\partial \Delta_{\mu}}{\partial x}\!\!\!\!\!&=&
-\rho^2\sum_{\alpha\beta}{\rm Tr}\Big(J^{\alpha}J^{\beta}\Big)\sum_{\nu\not=\mu}{\bf \Psi}_{\mu\nu}^{\alpha}{\bf \Psi}_{\nu\mu}^{\beta}\Delta_{\nu}\label{RG2},
\end{eqnarray}
where $x=\log (D/D_0)$, $D$ being the half of the scaled conduction-electron bandwidth, and $\rho$ is the density of states (DOS) at the Fermi level. We take the same bandwidth and the DOS for the all partial-wave components of the conduction electrons. The renormalization group equations, (\ref{RG1}) and (\ref{RG2}), are quite general and are not restricted in details of models\cite{comm}. In the following, we apply these general expressions to the four-level Kondo model.

\subsection{2-dimensional model}
In 2-dimensional configuration of Fig. {\ref{fig-1}}(a), we transform the ion operators $\phi_i$'s to the irreducible representations $a_{\mu}$'s of $D_{4h}$ point group as
\begin{eqnarray}
\Gamma_1^+:\ \ a_0&\equiv&\frac{1}{2}[\phi_1+\phi_2+\phi_3+\phi_4],\label{G1}\\
\Gamma_3^+:\ \ a_2&\equiv&\frac{1}{2}[\phi_1-\phi_2+\phi_3-\phi_4],\label{G3}\\
\Gamma_{5\pm}^-:\ \ a_{\pm1}&\equiv&\frac{1}{2}[\phi_1\mp i\phi_2-\phi_3\pm i\phi_{4}].\label{G5}
\end{eqnarray}
Furthermore, we need to define the operator form of ${\bf \hat{\Psi}}^{\gamma}$ (which is defined as $\hat{\bf \Psi}^{\gamma}\equiv a^{\dag}_{\mu}{\bf \Psi}^{\gamma}_{\mu\nu}a_{\nu}$) for the direct product $a^{\dag}_{\mu}a_{\nu}$:
\begin{eqnarray}
2\hat{\bf \Psi}^{0} &\equiv& a_0^{\dagger}a_{0}+a_{1}^{\dagger}a_1+ a_{2}^{\dagger}a_{2}+a_{-1}^{\dagger}a_{-1},\label{Ps0}\\
\sqrt{2}\hat{\bf \Psi}^{1} &\equiv& a_0^{\dagger}a_0-a_2^{\dagger}a_2,\\
\sqrt{2}\hat{\bf \Psi}^{2} &\equiv& a_1^{\dagger}a_1-a_{-1}^{\dagger}a_{-1},\\
2\hat{\bf \Psi}^{3} &\equiv& a_0^{\dagger}a_{0}-a_{1}^{\dagger}a_1+ a_{2}^{\dagger}a_{2}-a_{-1}^{\dagger}a_{-1},\\
2\hat{\bf \Psi}^{4} &\equiv& a_1^{\dagger}a_{0}+a_{2}^{\dagger}a_1+ a_{-1}^{\dagger}a_{2}+a_{0}^{\dagger}a_{-1},\\
2\hat{\bf \Psi}^{5}&\equiv& a_1^{\dagger}a_{0}-a_{2}^{\dagger}a_1- a_{-1}^{\dagger}a_{2}+a_{0}^{\dagger}a_{-1},\\
2\hat{\bf \Psi}^{6} &\equiv& -a_1^{\dagger}a_{0}+a_{2}^{\dagger}a_1- a_{-1}^{\dagger}a_{2}+a_{0}^{\dagger}a_{-1},\\
2\hat{\bf \Psi}^{7}&\equiv& -a_1^{\dagger}a_{0}-a_{2}^{\dagger}a_1+ a_{-1}^{\dagger}a_{2}+a_{0}^{\dagger}a_{-1},\\
\hat{\bf \Psi}^{8}&\equiv&[{  \hat{\bf \Psi}^4 }]^{\dag},\\
\hat{\bf \Psi}^{9}&\equiv&[  \hat{\bf \Psi}^5 ]^{\dag},\\
\hat{\bf \Psi}^{10}&\equiv&[ \hat{\bf \Psi}^6 ]^{\dag},\\
\hat{\bf \Psi}^{11}&\equiv&[ \hat{\bf \Psi}^7 ]^{\dag},\\
\sqrt{2}\hat{\bf \Psi}^{12} &\equiv& a_{1}^{\dagger}a_{-1}+a_{-1}^{\dagger}a_{1},\\
\sqrt{2}\hat{\bf \Psi}^{13} &\equiv& a_1^{\dagger}a_{-1}-a_{-1}^{\dagger}a_1,\\
\sqrt{2}\hat{\bf \Psi}^{14} &\equiv& a_0^{\dagger}a_{2}-a_{2}^{\dagger}a_0,\\
\sqrt{2}\hat{\bf \Psi}^{15} &\equiv& a_0^{\dagger}a_{2}+a_2^{\dagger}a_{0},\label{Ps15}
\end{eqnarray}
with constraint $2\hat{\bf \Psi}^0=\hat{\bf 1}$.
For the conduction electrons, we truncate the summation with respect to $l$ at $l=2$ in eq. (\ref{expLM}), because dominant contributions are expected to arise from the electrons with low $l$ waves. Corresponding irreducible representations for the quadratic form made from these conduction electron field operators are listed in Table {\ref{tbl-1}}.
\begin{table}[b]
	\begin{tabular}{|l|c|c|}
\hline
rep. $(D_{4h})$&conduction electron& ion \\
\hline
\hline
$\Gamma_1^+$ & $s^{\dag}s,$& ${\bf \Psi}^1,{\bf \Psi}^3$\\
&$d_0^{\dag}d_0,$&\\
&$s^{\dag}d_0+{\rm h.c.},$&\\
&$ d_+^{\dag}d_++d_-^{\dag}d_-,$&\\
&$d_+^{\dag}d_-+{\rm h.c.},$&\\
&$ p_+^{\dag}p_++p_-^{\dag}p_-$&\\
\hline
$\Gamma_2^+$ & $p_+^{\dag}p_+-p_-^{\dag}p_-,$ & ${\bf \Psi}^2$\\
&$ d_+^{\dag}d_+-d_-^{\dag}d_-,$&\\
&$d_+^{\dag}d_--{\rm h.c.}$&\\
\hline
$\Gamma_3^{(1)+}$ & $p_+^{\dag}p_-+{\rm h.c.},$ & ${\bf \Psi}^{12},{\bf \Psi}^{15}$\\
& $s^{\dag}(d_{+}+d_-)+{\rm h.c.},$ &\\
&$ d_0^{\dag}(d_{+}+d_-)+{\rm h.c.}$&\\
\hline
$\Gamma_3^{(2)+}$ & $s^{\dag}(d_{+}+d_-)-{\rm h.c.},$ & ${\bf \Psi}^{14}$\\
&$ d_0^{\dag}(d_{+}+d_-)-{\rm h.c.}$&\\
\hline
$\Gamma_4^+$ & $p_+^{\dag}p_--{\rm h.c.},$ & ${\bf \Psi}^{13}$\\
 & $ s^{\dag}(d_+-d_-)-{\rm h.c.},$ & \\
&$d_0^{\dag}(d_+-d_-)-{\rm h.c.}$&\\
\hline
$\Gamma_5^{(1)-}$ & $ \{ -s^{\dag}p_++p_-^{\dag}s, s^{\dag}p_--p_+^{\dag}s\}$,&$\{ {\bf \Psi}^{8}, {\bf \Psi}^{4}\},$\\
& $ \{ -d_0^{\dag}p_++p_-^{\dag}d_0, d_0^{\dag}p_--p_+^{\dag}d_0\}$,&
$\{ {\bf \Psi}^{9}, {\bf \Psi}^{5}\}$\\
 & $ \{ d^{\dag}_+p_--p_+^{\dag}d_-, -d_-^{\dag}p_++p_-^{\dag}d_+\}$,&\\
 & $ \{ d^{\dag}_-p_--p_+^{\dag}d_+, -d_+^{\dag}p_++p_-^{\dag}d_-\}$&\\
\hline
$\Gamma_5^{(2)-}$ & $ \{ -s^{\dag}p_+-p_-^{\dag}s, s^{\dag}p_-+p_+^{\dag}s\}, $ &$\{ -{\bf \Psi}^{10}, {\bf \Psi}^{6}\}$,\\
             & $ \{ -d_0^{\dag}p_+-p_-^{\dag}d_0, d_0^{\dag}p_-+p_+^{\dag}d_0\}, $ &$\{ -{\bf \Psi}^{11}, {\bf \Psi}^{7}\}$\\
 & $ \{ d^{\dag}_+p_-+p_+^{\dag}d_-, d_-^{\dag}p_++p_-^{\dag}d_+\},$&\\
 & $ \{ d^{\dag}_-p_-+p_+^{\dag}d_+, d_+^{\dag}p_++p_-^{\dag}d_-\}$&\\
\hline
        \end{tabular}
\caption{Irreducible representations for the quadratic terms of the conduction and the ion operators.
 We classify them by taking into account the fact that the couplings
 are all real quantities. In the second column, we use abbreviations such
 as $s=\sum_kc_{k00},\ p_{\pm}=\sum_kc_{k1\pm1},\ d_0=\sum_kc_{k20}$ and $d_{\pm}=\sum_kc_{k2\pm2}$.
 }
\label{tbl-1}
\end{table}

Then, the ``bare'' Hamiltonian is (see Appendix A for detailed derivation):

\begin{eqnarray}
H_{\rm int}\!\!\!\!\!&=& H_{{\rm diag}}+H_{\rm hop}, \label{Hhopdiag}\\
\frac{H_{{\rm diag}}}{u_0}\!\!\!\!\!&=& \frac{2k_Fa}{\sqrt{6}}\big \{s^{\dagger}p_{+}-p_{-}^{\dagger}s\big\}\hat{\bf \Psi}_{4}+\frac{2k_Fa}{\sqrt{6}}\big \{p_{+}^{\dagger}s-s^{\dagger}p_{-}\big\}\hat{\bf \Psi}_{8}\nonumber\\
\!\!\!\!\!&-&\!\!\!\!\!\frac{k_F^{2}a^2}{3}\big \{p_{+}^{\dagger}p_{-}+p_{-}^{\dagger}p_{+}-\sqrt{\frac{3}{10}}(s^{\dagger}(d_++d_-)+\ \rm h.c.\ )\big\}\nonumber\\
\!\!\!\!\!&\times&\!\!\!\!\!\frac{\hat{\bf \Psi}_{12}+\hat{\bf \Psi}_{15}}{\sqrt{2}},\label{Bare1-2D}
\end{eqnarray}

\begin{eqnarray}
\frac{H_{{\rm hop}}}{u_0e^{-a^2/(2\sigma^2)}}\!\!\!\!\!&=& \frac{\sqrt{2}k_F^{2}a^2}{6}\big\{ 2s^{\dagger}s +\frac{1}{\sqrt{5}} (s^{\dagger}d_0+d_0^{\dagger}s)\nonumber\\
\!\!\!\!\!&-&\!\!\!\!\!\sum_{\mu=\pm}p_{\mu}^{\dagger}p_{\mu}\big \}  \hat{\bf \Psi}_{1}+\frac{\sqrt{2}k_F^2a^2}{6}\big \{p_{+}^{\dagger}p_{-}-p_{-}^{\dagger}p_{+}\nonumber\\
\!\!\!\!\!&+&\!\!\!\!\!\sqrt{\frac{3}{10}}(s^{\dagger}(d_+-d_-)-\ \rm h.c.\ )\big\}\hat{\bf \Psi}_{13},\label{Bare2-2D}
\end{eqnarray}
 where we have used the abbreviations in Table \ref{tbl-1} for the conduction electron field operators, and keep only $l=0$ term in eq. (\ref{U}). In eq. (\ref{Hhopdiag}), there exist no $c_{k10}$ and $c_{k2\pm1}$ terms which emerge in the higher order interaction, {\it e.g.} in the case when we consider $u_1$. However, these terms are expected to be less relevant because of the 2-dimensional configurations of the ion (i.e., the ion is located on the xy-plane). For the local ion energy, we obtain:
\begin{eqnarray}
\sum_{\mu}\Delta_{\mu}a_{\mu}^{\dag}a_{\mu}=\Delta (a_0^{\dag}a_0-a_2^{\dag}a_2)\ \ \ \ \ \ (\Delta \le 0).\label{magfield}
\end{eqnarray}

\begin{figure}[t!]
	\begin{center}
    \includegraphics[width=.47\textwidth]{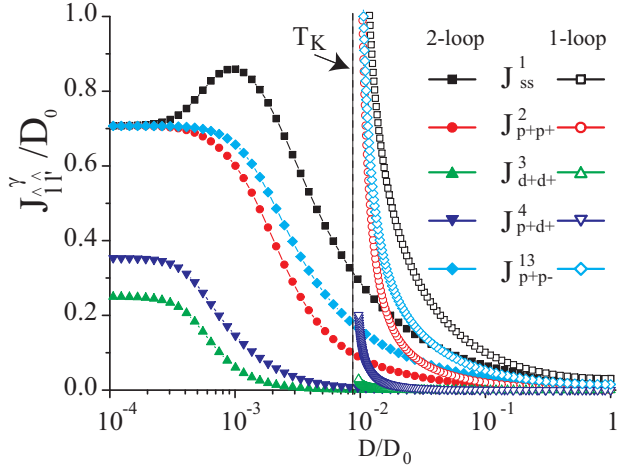}
  \end{center}
\caption{1- and 2-loop renormalization flow of the coupling $J^{\gamma}_{ij}$ (only typical couplings are shown). For example, a coupling $J_{p+p-}^{13}$ means one corresponding to the term $p_+^{\dag}p_{-}\hat{\bf \Psi}^{13}$. The Kondo temperature $T_K\simeq 0.009D_0$ of 1-loop result is shown by a vertical dotted line. The parameter set used is: $u_0=0.6D_0, k_Fa=0.8$, $k_F\sigma=0.42$, and $\Delta=-0.01D_0$. The values of $J^{\gamma}_{ij}$'s below $\sim T_K$ or $\tilde{\Delta}_{\mu}$ have no meaning because PRG would not be valid there.}
\label{2LRG}
\end{figure}

In Fig. \ref{2LRG}, we show the result of the 2-loop renormalization group flow for a typical case. It is noted that the 1-loop result is different from 2-loop one especially in $J_{p+d+}^{4}$ and $J_{d+d+}^{3}$. The fixed point of 2-loop calculation is an artifact arising from the third order approximation in $J_{\hat{l}\hat{l'}}^{\gamma}$ as in the single-channel Kondo effect\cite{Fowler,AbrMig}. Detailed NRG studies show that the fixed point given by a 1-loop result might be realized. At the fixed point of the 1-loop calculation (and of course around the same energy region for the 2-loop one), dominant coupling constants are those between $(p_+,s,p_-)$ and $(a_1,a_0,a_{-1})$, and become strong coupling.

 For $u_0>0$, the effective Hamiltonian can be written in the following form:
\begin{eqnarray}
H_{\rm int}^{\rm eff}\!\!\!\!&=&\!\!\!\!J_1(s_zS_z+\frac{1}{2}(q^{(2)}_+Q^{(2)}_-+q^{(2)}_-Q^{(2)}_+))+J_2(\frac{3}{2\sqrt{2}}q_zQ_z\nonumber\\
\!\!\!\!&&+\frac{1}{2}(s_+S_-+s_+S_++q^{(1)}_+Q^{(1)}_-+q^{(1)}_-Q^{(1)}_+)),\label{Heff2}
\end{eqnarray}
where $J_2\simeq \frac{5}{4}J_1\ge 0$, and $s_z(S_z)$ and $s_{\pm}(S_{\pm})$ are the z- and transverse-component of pseudospin $s(S)=1$ for the conduction electron (ion). The eigen states of $s_z$ and $S_z$ are constructed as 
\begin{eqnarray}
|s=1,s_z=0\rangle &=& s^{\dag}|0\rangle,\\
|s=1,s_z=\pm 1\rangle &=& \pm p_{\pm}^{\dag}|0\rangle,\\
|S=1,S_z=0\rangle &=& a_0^{\dag}|0\rangle,\\
|S=1,S_z=\pm 1\rangle &=& a_{\pm 1}^{\dag}|0\rangle.
\end{eqnarray}
On this basis set, $s$ and $q$ are expressed by $3\times 3$ matrices as follows
\begin{figure}[t!]
	\begin{center}
    \includegraphics[width=.47\textwidth]{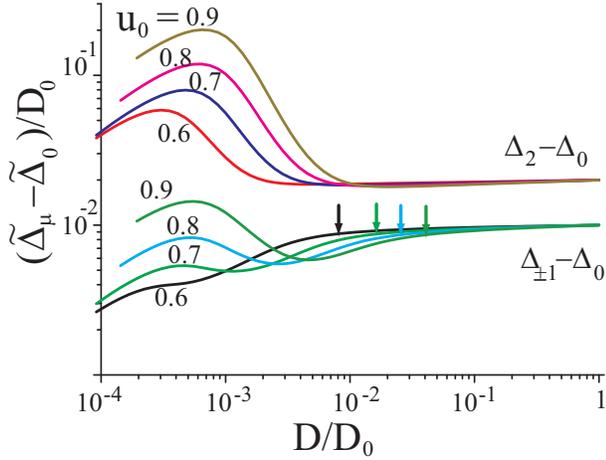}
  \end{center}
\caption{Renormalization evolution of the energy splitting of the ion configurations $\tilde{\Delta}_{\mu}$ as a function of the scaled bandwidth $D/D_0$. Arrows indicate the 
Kondo temperature $T_K$ given by the 1-loop calculation. The calculated $\tilde{\Delta}_{\mu}$'s have no meaning below $\sim T_K$ or $\tilde{\Delta}_{\mu}$.}
\label{fig-3}
\end{figure}
\begin{figure}[h!]
\begin{center}
    \includegraphics[width=.47\textwidth]{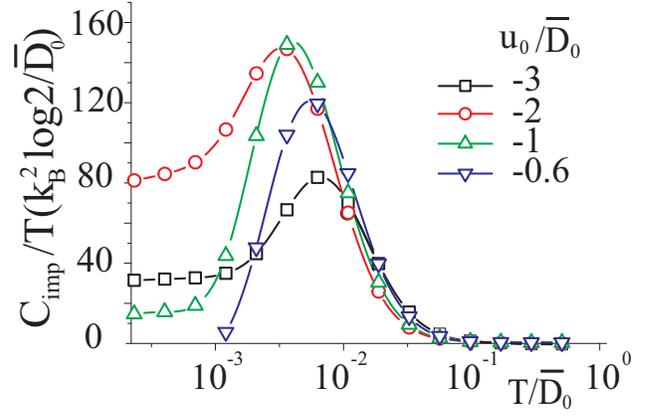}
\end{center}
\caption{$C_{{\rm imp}}/T$ vs. $T$, $C$ being the impurity specific heat. The parameter set used is: $\Delta=-0.0169\bar{D}_0$, $k_Fa=0.8$ and $k_F\sigma=0.42$ (we obtain similar results for $u_0>0$ cases). For the bare bandwidth $\bar{D}_0/k_{\rm B}=100$K, $C_{\rm imp}/T\simeq 57\times {\rm (scale\ of\ ordinate)\ mJ/{\rm K^2}}\cdot \rm mol$.}
\label{fig-4}
\end{figure}
\begin{figure}[h!]
\begin{center}
    \includegraphics[width=.47\textwidth]{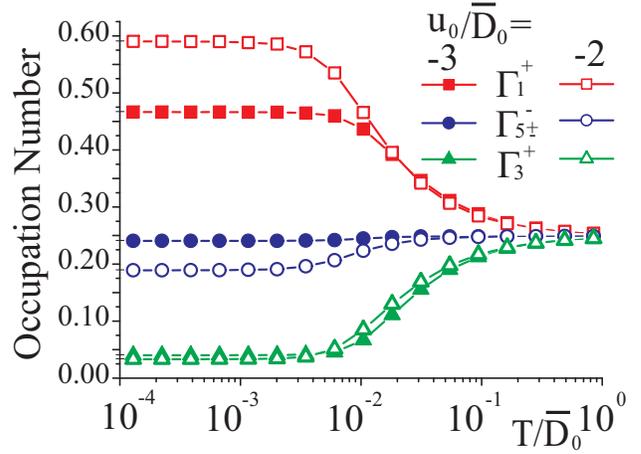}
\end{center}
\caption{Occupation number of ionic configurations $\Gamma's$ as a function of temperature. The parameters used are the same as those in Fig. \ref{fig-4}. Although we show results for $u_0<0$ cases, we obtain similar results for $u_0>0$ cases.}
\label{fig-5}
\end{figure}
\begin{figure}[h!t]
\begin{center}
    \includegraphics[width=.43\textwidth]{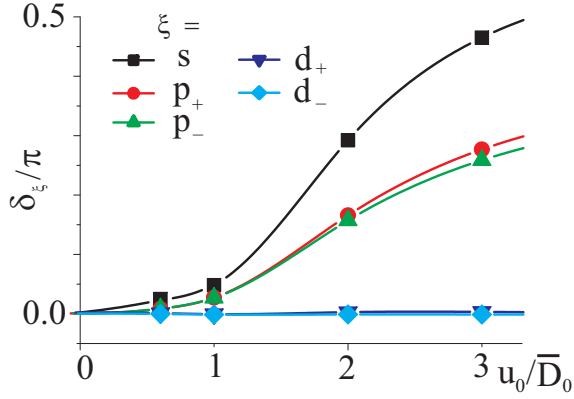}
\end{center}
\caption{Phase shifts $\delta_{\xi}$ of the conduction electrons with $\xi$-symmetry vs. the interaction $u_0$. The parameters used are the same as those in Fig. \ref{fig-4}. A similar result is obtained for $u_0<0$. The small splitting between $p_+$ and $p_{-}$ at $u_0=3\bar{D}_0$ is due to a numerical error from truncation procedures in NRG.}
\label{fig-6}
\end{figure}
\begin{eqnarray}
s_z: \left(
\begin{array}{@{\,}ccc@{\,}}
   1 & 0 & 0 \\
   0 & 0 & 0 \\
   0 & 0 & -1 \\
\end{array}
\right),\  
s_+: \left(
\begin{array}{@{\,}ccc@{\,}}
   0 & \sqrt{2} & 0 \\
   0 & 0 & \sqrt{2} \\
   0 & 0 & 0 \\
\end{array}
\right),
\end{eqnarray}

\begin{eqnarray}
q_z: \left(
\begin{array}{@{\,}ccc@{\,}}
   \frac{-1}{\sqrt{3}} & 0 & 0 \\
   0 & \frac{2}{\sqrt{3}} & 0 \\
   0 & 0 & \frac{-1}{\sqrt{3}} \\
\end{array}
\right),\ 
q_+^{(1)}: \left(
\begin{array}{@{\,}ccc@{\,}}
   0 & \sqrt{2} & 0 \\
   0 & 0 & -\sqrt{2} \\
   0 & 0 & 0 \\
\end{array}
\right),
\end{eqnarray}
\begin{eqnarray}
q_+^{(2)}: \left(
\begin{array}{@{\,}ccc@{\,}}
   0 & 0 & 2 \\
   0 & 0 & 0 \\
   0 & 0 & 0 \\
\end{array}
\right),
\end{eqnarray}
 and $s_{-}=s_+^{\dag}$, $q_-^{(1)}={q_+^{(1)}}^{\dag}$ and $q_-^{(2)}={q_+^{(2)}}^{\dag}$. The operators $q$'s represent the quadrupolar degrees of freedom. $S_z, Q_z$, etc. are defined in the same way. In these representations the ion energy of eq. (\ref{magfield}) implies a local distortion field in pseudospin space, because the following relation holds:
\begin{eqnarray}
a_0^{\dag}a_0-a_2^{\dag}a_2\simeq a_0^{\dag}a_0=\frac{\sqrt{3}Q_z+\hat{\bf 1}}{2},
\end{eqnarray}
near the strong coupling fixed point where $a_2^{\dag}a_2$ is irrelevant (the $\Gamma_3^+$ degrees of freedom are frozen out in low temperature).

 For $u_0<0$, we also obtain an expression for the effective Hamiltonian similar to eq. (\ref{Heff2}). From eq. (\ref{Heff2}), it is expected that an anisotropic spin 1 dipolar $(s_z, s_{\pm})$ and quadrupolar $(q_z, q_{\pm}^{(1)}, q_{\pm}^{(2)})$ Kondo effect with the conduction electron with the pseudospin 1 occurs in this model. For comparison, the more symmetric Kondo model, in which the impurity and the conduction electron with spin 1 are coupled isotropically, has two fixed points. One is the SU(3) symmetric fixed point, which is occurred when we take into account both the dipolar and the quadrupolar interactions. The other is a non-Fermi liquid one keeping the only dipolar interaction.\cite{Hat}

In Fig. \ref{fig-3}, the energy levels of the various ion configurations are shown. A renormalization of $\Delta_{\mu}$ does not occur at the level of 1-loop PRG (poorman's scaling) but appears at the 2-loop PRG equation eq. (\ref{RG2}). It is noted that the energy level $\Delta_{\pm 1}$ of the first excited state $\Gamma_{5}^{-}$ decreases as the effective conduction electron bandwidth decreases. On the other hand, that of the second excited state $\Gamma_3^{+}$, $\Delta_2$,  increases. The results in the low $D/D_0$ region are unphysical ones, because the perturbative approach cannot be reliable there (especially $D<{\rm max}(|\tilde{\Delta}_{\mu}-\tilde{\Delta}_{0}|,T_K)$).

\setcounter{equation}{0}
\section{NRG Results}
In this section, we investigate the 2-dimensional model (eq. (\ref{Hhopdiag})) using the NRG method\cite{Wilson}. Because of computational restrictions, we take the only $s$, $p_{\pm}$ and $d_{\pm}$ conduction electrons into account. The omitted orbital $d_0$ is expected to play an irrelevant role from the result due to the PRG study discussed in the previous section.
For the NRG calculations, it is helpful to introduce the following two conserved quantities:

\begin{eqnarray}
Q&\equiv&\sum_{klm}(c_{klm}^{\dagger}c_{klm}-1/2),\\
h&\equiv&\!\!\!\!\!{\rm mod}(\sum_{klm}mc_{klm}^{\dagger}c_{klm}+\!\!\!\!\!\sum_{\mu=\pm 1,0,2}\mu a_{\mu}^{\dagger}a_{\mu},\ 4),
\end{eqnarray}
where $Q$ is the total charge of the conduction electrons measured from the half filling and the {\it helicity} $h$ is a kind of the z-component of the angular momentum. In the NRG calculation, we transform the conduction-electron operators $c_{klm}$ into the 1-dimensional semi-infinite chains $f_{nlm}\ \ (n=0,1,2,...)$ and diagonalize iteratively the Hamiltonian starting from the $n=0$ shell. The conduction-electron part in the Hamiltonian (\ref{Heff}) is expressed as follows:
\begin{eqnarray}
\!\!\!\!\!\!\!\!\!\!\sum_{klm}\epsilon_k c_{klm}^{\dag}c_{klm}\!\!\!\!&=&\!\!\!\!\!\sum_{n=0}^{\infty}\sum_{lm}t_n\Lambda^{-\frac{n}{2}}f_{nlm}^{\dag}f_{n+1lm}+\rm h.c.,\\
t_n&=&\frac{(1+\Lambda^{-1})(1-\Lambda^{-n-1})}%
          {2\sqrt{(1-\Lambda^{-2n-1})(1-\Lambda^{-2n-3})}},
\end{eqnarray}
 where $\Lambda$ is a logarithmic discretization parameter in the NRG calculation. We keep 1200 states at each iteration and use $\bar{D}_0\equiv 2D_0/(1+\Lambda^{-1})$ as a unit of the energy and set $\Lambda=3$ in this section.

 In Fig. \ref{fig-4}, we show the result of $C_{\rm imp}/T$ the impurity specific heat (divided by temperature). Although we show results for $u_0<0$ case in Fig. {\ref{fig-4}}, we obtain similar results for $u_0>0$ case. In the region of small $u_0$, we do not have a heavy effective mass, because the $T_K$ in this region cannot exceed the local splitting $\Delta$. In the region of large $u_0$, say for $u_0/\bar{D}_0=-3$ in Fig. \ref{fig-4}, the $T_K$ can overwhelm the splitting $\Delta$. In this case, however, the $T_K$ is too large so that the effective mass becomes small. As a result, we have the heaviest effective mass at an intermediate strength of $u_0$ i.e., $u_0/\bar{D}_0=-2$ in Fig.\ref{fig-4}. These features are common in the usual Kondo model under a magnetic field, which has an equivalent function to the local splitting in the present model. A size of $C_{\rm imp}/T$ in a physical unit depends on how to take $D_0 (\bar{D}_0)$. 
If we set $\bar{D}_0/k_{\rm B}=100$K as discussed at the beginning of \S\ \ref{3}, it is given as $C_{\rm imp}/T\simeq 57\ \times $ (scale of ordinate) mJ/(K$^2\cdot$mol). Then for $u_0=-2\bar{D}_0$, $\lim_{T\to 0}C_{\rm imp}/T\simeq 4.6\times 10^{3}$ mJ/(K$^2\cdot$mol), which implies that local heavy electron state is realized.

The occupation number for each configuration of the ion is shown in Fig. \ref{fig-5}. As expected from the results in the last section, a Kondo effect between $(a_{1},a_0,a_{-1})$ and $(p_+,s,p_-)$ seems to be occurred. It is clearly seen in Fig. \ref{fig-5} that the orbitals, which take part in the Kondo effect, grows at low temperature. It is noted that the state $\Gamma_{5\pm}^-$, originally doublet excited states, are included in the manybody ground state together with $\Gamma_1^+$ fully symmetric state. This is marked contrast with a simple picture of quantum tunneling among multistable points and also with the case of two-level Kondo problem in which an ionic configuration of the ground state is fully symmetric like a simple TLS.

 In Fig. \ref{fig-6}, we show the phase shift of conduction electrons $\delta_{\xi}\ (\xi=s,p_{\pm},d_{\pm})$. We estimate $\delta_{\xi}$ using the energy level of the first excited states in each $\xi$. It is noted that $\delta_{d_{\pm}}\simeq 0$ for any values of $u_0$ while $\delta_{\xi}\ (\xi=s,p_{\pm})$ are enhanced as $u_0$ is increased. This confirms the result by the PRG approach discussed in \S\ 3, where it has been shown that $d_{\pm}$ orbitals do not take part in the fixed point Hamiltonian and are decoupled.

\section{Discussions}
 As discussed above, a realization of Kondo screening in multilevel systems crucially depends on the strength of the coupling constant between the conduction electrons and the ion charge fluctuations. 
In the last section, we have used the coupling $u_0=O(1)\times D_0$. It appears that such a coupling is unrealistically large. However, we have to remember that $D_0$ is an effective bandwidth of the order $\hbar\omega_0$. As argued by Aleiner {\it et al}.\cite{Aleiner}, the conduction electrons with energy $\gg \hbar\omega_0$ may not contribute a renormalization of the coupling $u_{l}$ in the present model when the effective bandwidth is scaled down. As a result, we can use relatively strong coupling constants and splittings of ion's level. 
 Since the spin degrees of freedom have been neglected in the present model, the enhanced effective mass obtained here is evidently robust against a magnetic field. It is interesting to note that the heavy electron state, which is robust against the magnetic field up to 14T, has been observed in SmOs$_4$Sb$_{12}$\cite{Sm}.

Although the present model (2-dimensional version) cannot be compared directly to the real materials, the important result is that {\it it is possible to include anisotropic ion modes in the ground state wavefunction if the coupling between conduction electrons and the ion charge fluctuation is moderately strong}. This ``inclusion'' is expected even at the level of 1-loop renormalization group. Indeed, since the effective Hamiltonian at the fixed point is written as an anisotropic spin-1 Kondo Hamiltonian, the selected ion's modes $\Gamma_1^+$ and $\Gamma_{5\pm}^-$ have energy gains by the Kondo effect. 
In 3-dimensional model (Fig.\ref{fig-1}(b)), a similar renormalization is expected to occur, involving the anisotropic $\Gamma_4$ triplet configuration (excited state) together with the isotropic $\Gamma_1$ configuration.

 In order to discuss the thermal rattling, it becomes necessary to include more realistic aspects in the model, {\it e.g.}, the role of excited states of the ionic motion and the effect of dissipation in the local system. This is because, in ultrasonic experiments, the frequency $\omega$ of the ultrasonic is $\omega\sim$ 1mK, so that it is difficult to realize the dispersion of elastic constants at around 30K with respect to the frequency\cite{NemotoCePd, RevClath,GotoPrOs} only by taking into account the complete energy levels of the local system without dissipation. An important ingredient giving the dispersion would be the dissipation due to the interaction between the electron and the local ion, or that between the local ion and the acoustic phonons. Indeed, in refs. 3-5, a Debye-type dispersion is used to analyze the results of the experiments.
The reason why the ultrasonic dispersion (rattling) is observed only in specific modes might be related to the matrix elements of the dissipative interaction.

It is noted that there are two aspects in this problem. One is the physics of the heavy fermion in which the several lowest-energy levels of the local system and the orbital Kondo effect play an important role as discussed in this paper. The other is that of the thermal excitations coupled to the time-dependent external perturbation such as the ultrasonic, which cause so-called rattling.

For treating excited states, it is needed to solve the 3-dimensional Schr\"odinger equation directly, assuming a proper off-center potential. Recently, a band calculation was performed modeling $\beta$-pyrochlore compounds\cite{KOsO}. The eigen value of the energy and the wavefunction for low-lying ionic states were calculated, although the couplings between the conduction electrons and the ion have not been estimated, because the estimation of the coupling constants is highly nontrivial.

In the filled-skutterudite compounds, anomalous phonon contributions are observed in various quantities, especially in ROs$_4$Sb$_{12}$ (R=Pr, La, Sm, etc.). One of the reasons why the effect is prominant in ROs$_4$Sb$_{12}$ may be that the size of Sb$_{12}$ cage is the largest among existing filled-skutterudite compounds including rare earth ion R. Another one may be that the conduction electrons near the Fermi level in Os-compounds consist of a molecular orbital of the cage (A$_{1u}$) and the d-electron in Os. Unlike Ru-compound, in which the levels of 4d electrons are deep compared to that of 5d or 3d electrons so that their components near the Fermi level are negligible, the contributions from d-electrons are much larger. Then it is expected that d-electrons play important roles in Os compounds. 
 For PrOs$_4$Sb$_{12}$, it might be speculated that something related to the off-center degrees of freedom take an important role for the superconducting transition. But the relation between these local phonons and unconventional superconductivity is beyond a scope of the present paper, and further studies are required.


\section{Conclusions}
 In summary, we have studied the four-level system with spinless conduction electrons by using renormalization group methods. The relevant component of the ion configurations are $\Gamma_1^+$ and $\Gamma_5^-$ which couple to the $s$ and $p_{\pm}$ conduction electrons in the low temperature fixed point. The low energy effective Hamiltonian has been written in a pseudospin-1 language both for the ion and the conduction electrons with a pseudo-distortion field. As a result, anisotropic components of the ion off-center modes enter the ground state wavefunction. We expect that the observed relatively large effective mass and the robustness to the magnetic field in SmOs$_4$Sb$_{12}$ and other heavy-fermion-like materials without f-electrons can be explained.
\section*{Acknowledgment}

We would like to thank Y. Nemoto and T. Goto for fruitful discussions. One of us (K. H.) is supported by Research Fellowships of JSPS for Young Scientists.
This work is supported by a Grant-in-Aid for Scientific Research (No.16340103), 21st Century COE Program (G18) by Japan Society for the Promotion of Science, and a Grant-in-Aid for Scientific Research in Priority Areas (No. 16037209) by MEXT.

\appendix
\section{A: Derivation of effective Hamiltonian for $D_{4h}$ symmetry}
 In this appendix, we derives the bare Hamiltonian eqs. (\ref{Bare1-2D}) and (\ref{Bare2-2D}). First of all, Hamiltonian eq. (\ref{Hint}) is transformed into the form of eq. (\ref{HintKondo}). 
\begin{eqnarray}
\!\!\!\!\!\!\!\!\!\! H_{\rm int}\!\!\!\!\!&=&\!\!\!\!\!\!\sum_k\sum_{k'}\!\!\sum_{lm}\sum_{l'm'}\sum_{ij}c^{\dag}_{klm}c_{k'l'm'}\phi_i^{\dag}\phi_j\nonumber\\
\!\!\!\!\!\!\!\!\!\!\!\!\!\!&\times&\!\!\!\!\!\!\!\Bigg[{\rm i}^{(l'-l)}\!\!\int\!\!\!\!\int \!\!\frac{\Omega_k}{4\pi}\frac{\Omega_{k'}}{4\pi}Y_{lm}^*(\hat{k})U_{\bf q}F_{ij}({\bf q})Y_{l'm'}(\hat{k'})\Bigg].\label{A2}
\end{eqnarray}
where ${\bf q=(k-k')}/k_F$ and notice that $\sum_k$ involves only a summation of the radial part $k$.

  $F_{ij}({\bf q})$ for $i\not= j$ in eqs. (\ref{Hint}) and (\ref{A2}) is approximated as
\begin{eqnarray}
\!\!\!\!\!\!\!\!\!\!\!\!\!\!\!\!\!\!\!\!\!\!\!\!F_{ij}({\bf q})&\simeq& \exp(-\frac{a^2}{2\sigma^2})\frac{(k_F a)^2}{4}\nonumber
\end{eqnarray}
\begin{eqnarray}
\times\left \{
\begin{array}{@{\,}c@{\,}}
   -q_xq_y+\frac{1}{2}(q_x^2+q_y^2) \ \ \ {\rm for}\ (i,j)=(1,2),\ (3,4)\\
   +q_xq_y+\frac{1}{2}(q_x^2+q_y^2) \ \ \ {\rm for}\ (i,j)=(2,3),\ (4,1)\\
\end{array}
\right \},
\end{eqnarray}
where we have expanded $F_{ij}({\bf q})$ with respect to $k_Fa(\ll 1)$ and approximated $\exp(-\frac{\sigma a^2|{\bf q}|^2}{4})\simeq 1$. In the same way, for $i=j$, we obtain 
\begin{eqnarray}
\!\!\!\!\!\!\!\!\!\!\!\!\!\!\!\!\!\!\!\!\!\!\!\!F_{ij}({\bf q})\simeq\nonumber
\end{eqnarray}
\begin{eqnarray}
\left \{
\begin{array}{@{\,}c@{\,}}
   1-{\rm i}(k_F a)q_x-\frac{(k_F a)^2}{2}q_x^2 \ \ \ {\rm for}\ (i,j)=(1,1)\\
   1-{\rm i}(k_F a)q_y-\frac{(k_F a)^2}{2}q_y^2 \ \ \ {\rm for}\ (i,j)=(2,2)\\
   1+{\rm i}(k_F a)q_x-\frac{(k_F a)^2}{2}q_x^2 \ \ \ {\rm for}\ (i,j)=(3,3)\\
   1+{\rm i}(k_F a)q_y-\frac{(k_F a)^2}{2}q_y^2 \ \ \ {\rm for}\ (i,j)=(4,4)\\
\end{array}
\right \}.\label{A4}
\end{eqnarray}
In the case of only retaining $u_0$ for $U_{\bf q}$, the next task is to estimate matrix elements of $(q_x,q_y,q_x^2,q_xq_y,q_y^2)$ in the square brackets of eq. (\ref{A2}). For the calculation of these matrix elements, it is useful to introduce the following spherical harmonics representations:
\begin{eqnarray}
\left \{
\begin{array}{@{\,}c@{\,}}
x\\
-{\rm i}y
\end{array}
\right \}
&=&\sqrt{\frac{2\pi}{3}}(Y_{1-1}\mp Y_{11}),\\
\left \{
\begin{array}{@{\,}c@{\,}}
x^2\\
y^2
\end{array}
\right \}&=&\!\!\!\sqrt{\pi}\Bigg[ \frac{2}{3}Y_{00}-\frac{2}{3\sqrt{5}}Y_{20}\pm\sqrt{\frac{2}{15}}(Y_{22}+Y_{2-2})   \Bigg],\nonumber\\
\\
-{\rm i}xy&=&\sqrt{\frac{2\pi}{15}}(Y_{2-2}-Y_{22}).
\end{eqnarray}
As the matrix elements of 
\begin{eqnarray}
{\rm i}^{(l'-l)}\int\!\!\!\int\frac{d\Omega_k}{4\pi}\frac{d\Omega_{k'}}{4\pi} Y^*_{l,m}(\hat{k})(q_x,q_y,q_x^2,q_xq_y,q_y^2)Y_{l',m'}(\hat{k}'),\nonumber
\end{eqnarray}
we obtain the following results,
\begin{eqnarray}
q_x&:&\frac{-{\rm i}}{4\pi\sqrt{6}}\delta_{l1}\delta_{l'0}(\delta_{m-1}-\delta_{m1})+\rm perm.,\\
q_y&:&\frac{1}{4\pi\sqrt{6}}\delta_{l1}\delta_{l'0}(\delta_{m-1}+\delta_{m1})+\rm perm.,\\
\left \{
\begin{array}{@{\,}c@{\,}}
q_x^2\\
q_y^2
\end{array}
\right \}\!\!\!\!\!&:&\!\!\!\!\!\Bigg[\frac{\delta_{l'0}}{12\pi}\Big( \delta_{l0} +\frac{\delta_{l2}}{\sqrt{5}}[\delta_{m0}\mp \sqrt{\frac{3}{2}}(\delta_{m2}+\delta_{m-2})]  \Big)\nonumber\\
\!\!\!\!\!&+&\!\!\!\!\!\rm perm.\Bigg]\nonumber\\
\!\!\!\!\!&-&\!\!\!\!\!\frac{1}{12\pi}\delta_{ll'}\delta_{l1}(\delta_{m-1}\mp \delta_{m1})(\delta_{m'-1}\mp\delta_{m'1}),\\
q_xq_y&:&\Bigg[\frac{-{\rm i}}{4\pi\sqrt{30}}\delta_{l'0}\delta_{l2}(\delta_{m-2}-\delta_{m2})-{\rm perm.}\Bigg]\nonumber\\
&-&\frac{{\rm i}}{12\pi}\delta_{ll'}\delta_{l1}(\delta_{m1}\delta_{m'-1}-\delta_{m-1}\delta_{m'1}).
\end{eqnarray}
Here, ``perm.'' means $lm \leftrightarrow l'm'$.
Finally, we use eqs. (\ref{G1})-(\ref{Ps15}), and neglect simple potential scattering terms corresponding to the first term in eq. (\ref{A4}), then we obtain eqs. (\ref{Bare1-2D}) and (\ref{Bare2-2D}).


\vspace{1cm}
{\bf Note added:}\\

Recently, Matsuhira et al. [19aYG-13, JPS topical meeting, 19-22 September, Tanabe, Japan] reported a result of specific heat measurements in
LaOs$_4$P$_{12}$. They find that an anomalous increase in the specific heat ($C/T\propto T^{-2}$) at low temperatures ($T\le 2$K). Surprisingly, this behavior is robust against magnetic fields so that they claimed it is an intrinsic effect other than the nuclear spin contribution. We believe that this power-law behavior is explained by the peak structure of the specific heat in Kondo systems. Although it is not well recognized, $C_{\rm imp}/T$ in the high temperature side of this peak is well described by $T^{-2}$ law. Indeed, the result of $C_{\rm imp}/T$ by NRG calculation exhibits such $T$-dependence as shown in the figure below.
From our point view, the anomalous temperature dependence might arise from a characteristic of this system in which the very low Kondo temperature of the ionic off-center motions is realized due to a relatively weak-coupling between multilevel system and conduction electrons in P$_{12}$ cage with smaller sized compared to Sb$_{12}$ cage. 
\vspace{1cm}
\begin{figure}[h!]
\begin{center}
    \includegraphics[width=.45\textwidth]{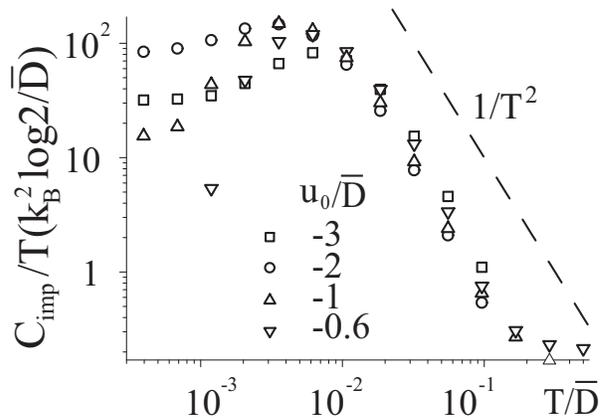}
\end{center}
\caption{$C_{{\rm imp}}/T$ vs $T$. The parameter set used is as same as that in Fig. 4.}
\end{figure}


\begin{thebibliography}{99} 
\bibitem{Kondo1}J. Kondo: Physica B+C {\bf 84} (1976) 40.
\bibitem{Kondo2} J. Kondo: Physica B+C {\bf 84} (1976) 207.
\bibitem{NemotoCePd} Y. Nemoto, T. Yamaguchi, T. Horino, M. Akatsu, T. Yanagisawa, T. Goto, O. Suzuki, A. D$\rm \ddot{o}$nni and T. Komatsubara: Phys. Rev. B {\bf 68} (2003) 184109.
\bibitem{RevClath} T. Goto, Y. Nemoto, T. Yamaguchi, M. Akatsu, T. Yanagisawa, O. Suzuki and H. Kitazawwa: Phys. Rev. B {\bf 70} (2004) 184126.

\bibitem{GotoPrOs} T. Goto, Y. Nemoto, K. Sakai, T. Yamaguchi, M. Akatsu, T. Yanagisawa, H. Hazama, K. Onuki, H. Sugawara and H. Sato: Phys. Rev. B {\bf 69} (2004) 180511R.
\bibitem{NaCl} S. Suto and M. Ikezawa: J. Phys. Soc. Jpn. {\bf 53} (1984) 438.
\bibitem{NaCl2} H. Yamada, S. Tanaka, Y. Kayanuma and T. Kojima: J. Phys. Soc. Jpn {\bf 54} (1985) 1180.
\bibitem {Sm} S. Sanada, Y. Aoki, H. Aoki, A. Tsuchiya, D. Kikuchi, H. Sugawara and H. Sato: J. Phys. Soc. Jpn {\bf 74} (2005) 246.
\bibitem{YuAnderson} C. Yu and P. W. Anderson: Phys. Rev. B {\bf 29} (1984) 6165.
\bibitem{MatsuMk} T. Matsuura and K. Miyake: J. Phys. Soc. Jpn. {\bf 55} (1986) 610.
\bibitem {KusuMyk} H. Kusunose and K. Miyake: J. Phys. Soc. Jpn. {\bf 65} (1996) 3032.
\bibitem {MikimotoOno} K. Mitsumoto and Y. \=Ono: Physica C {\bf 426-431} (2005) 330.
\bibitem {VladZw1} K. Vlad$\rm \acute{a}$r and A. Zawadowski: Phys. Rev. B {\bf 28} (1983) 1564.
\bibitem {VladZw2} K. Vlad$\rm \acute{a}$r and A. Zawadowski: Phys. Rev. B {\bf 28} (1983) 1582.
\bibitem {VladZw3} K. Vlad$\rm \acute{a}$r and A. Zawadowski: Phys. Rev. B {\bf 28} (1983) 1596.
\bibitem{ZarandZwadow} G. Zar$\rm \acute{a}$nd and A. Zawadowski: Phys. Rev. Lett. {\bf 72} (1994) 542.
\bibitem{Moustakas1} A. L. Moustakas and D. S. Fisher: Phys. Rev. B {\bf 51} (1995) 6908.
\bibitem{Moustakas2} A. L. Moustakas and D. S. Fisher: Phys. Rev. B {\bf 55} (1997) 6832.
\bibitem{Ye} J. Ye: Phys. Rev. B {\bf 56} (1997) 1316.
\bibitem {MLS} G. Zar\'and: Phys. Rev. Lett. {\bf 77} (1996) 3609.
\bibitem{Wilson} K. G. Wilson: Rev. Mod. Phys. {\bf 47} (1975) 773.
\bibitem {CoxZW} D. L. Cox and A. Zawadowski: {\it Exotic Kondo Effects in Metals} (Taylor \& Francis, London, 1999), Appendix A.
\bibitem{Aleiner} I. L. Aleiner, B. L. Altshuler, Y. M. Galperin and T. A. Shutenko: Phys. Rev. Lett. {\bf 86} (2001) 2629.
\bibitem{resonant} G. Zar\'and: condmat/0411348.
\bibitem{KusuKura} H. Kusunose and Y. Kuramoto: Phys. Rev. B {\bf 59} (1999) 1902.
\bibitem{Yotsu} S. Yotsuhashi, M. Kojima, H. Kusunose and K. Miyake: J. Phys. Soc. Jpn. {\bf 74} (2005) 49.
\bibitem{comm} Eq. (\ref{RG2}) is applicable to the systems, which have no degeneracy in the same irreducible representation of the ion. It seems sufficient to use eq. (\ref{RG2}) when we consider the system in which there are not so many stable points of the ion's off-center motion.
\bibitem{Fowler} M. Fowler and A. Zawadowski: Solid State Commun. {\bf 9} (1971) 471.
\bibitem{AbrMig} A. A. Abrikosov and A. A. Migdal: J. Low Temp. Phys. {\bf 3} (1970) 519.
\bibitem{Hat} K. Hattori: unpublished.
\bibitem{KOsO} J. Kune\v{s}, T. Jeong and W. E. Pickett: Phys. Rev. B {\bf 70} (2004) 174510.
\end{thebibliography}
\end{document}